\documentclass{llncs}
\usepackage{graphicx}
\usepackage{url}
\usepackage{common}
\usepackage{footmisc}
\usepackage{subcaption}
\usepackage{array}
\usepackage{ragged2e}
\newcolumntype{P}[1]{>{\RaggedLeft\hspace{0pt}}p{#1}}
\title{MeetupNet Dublin: Discovering Communities in Dublin's Meetup Network\thanks{This research was supported by Science Foundation Ireland (SFI) under Grant Number SFI/12/RC/2289.}}
\author{Arjun Pakrashi, Elham Alghamdi, Brian Mac Namee, Derek Greene}
\institute{University College Dublin, Ireland \\
\email{arjun.pakrashi@ucdconnect.ie, elham.alghamdi@ucdconnect.ie, derek.greene@ucd.ie, brian.macnamee@ucd.ie}
}

\begin{document}
\maketitle

\begin{abstract}

Meetup.com is a global online platform which facilitates the organisation of meetups in different parts of the world. A meetup group typically focuses on one specific topic of interest, such as sports, music, language, or technology. However, many users of this platform attend multiple meetups. On this basis, we can construct a co-membership network for a given location. This network encodes how pairs of meetups are connected to one another via common members. In this work we demonstrate that, by applying techniques from social network analysis to this type of representation, we can reveal the underlying meetup community structure, which is not immediately apparent from the platform's website. Specifically, we map the landscape of Dublin's meetup communities, to explore the interests and activities of meetup.com users in the city.

\end{abstract}

\section{Introduction}

Meetup.com is an online platform that helps people to plan, organise, and discover public or private events, referred to as \textit{meetups}. Members of the platform can join specific meetup groups, that are usually based around a specific topic or activity, within which meetup events are organised. Meetup.com hosts groups that focus on diverse topics including sports, food, language, technology, business, philosophy, and dancing. The meetup.com platform is used worldwide and at the time of writing hosted more than 300k meetup groups and 39m users. On average over 3m people attend a meetup event each month. 

A complex network structure underlies the meetup.com platform, consisting of connections between users, meetup groups, and meetup events. By applying popular techniques from the field of social network analysis, such as centrality analysis \cite{bonacich1987power} or community finding \cite{girvan2002community}, we can potentially reveal insights about a given city or country. This could range from the most active sports communities, to the most popular types of music, to the most vibrant tech scenes. These insights are potentially useful to meetup.com organisers and members, recruiters, entrepreneurs, tourists, and city planners.

In this paper, we describe an analysis of the meetup.com network in Dublin, Ireland. Rather than focusing on individual users of the platform, we instead form a network representation at the meetup level, where a connection between two meetups exists if the two groups share members in common. We focus on two key research questions: 1) do distinct thematically-coherent communities exist within Dublin's Meetup ecosphere?; 2) if so, how do these communities overlap with one another? 
To answer these questions, we apply the popular \emph{OSLOM} (Order Statistics Local Optimisation Method) algorithm \cite{lancichinetti11oslom} to discover overlapping communities of connected meetups, and text analytics methods to summarise these communities. This analysis reveals Dublin's place as a technology capital through a large number of communities of large technology focused meetups, the importance of meetups focusing on generic topics such as language learning as connectors within the community, and how large communities of meetups are linked by a small number of meetups. This analysis is a demonstration of how the meetup.com network structure can be used to gain insights into the communities that exist within a city and how they interact. To support the further analysis of meetup groups in other geographic locations, we make the relevant code and data available for reuse\footnote{\label{note1}Code: \url{https://github.com/phoxis/MeetupNetDublin}\\Interactive graph: \url{https://draig.ucd.ie/MeetupNetDublinInteractive/}}.

The remainder of this paper proceeds as follows. \refsec{sec:related} describes related work on co-occurrence networks of the type used in this paper and community finding. \refsec{sec:methods} describes how the meetup network was constructed, how community finding techniques were applied to it, and how the resulting communities were labelled using textual metadata. \refsec{sec:eval} explores the network created and the communities found within it, focusing on a subset of largely technology-focused communities. Finally, \refsec{sec:conclusions} concludes with suggested directions for future work in this area.


\section{Related Work}
\label{sec:related}


\subsection{Analysing Co-occurrence Networks} 

A \emph{co-occurrence network} is a weighted network constructed such that each edge indicates the frequency with which two items appear in the same context. In some cases, this kind of network is formed by projecting an existing bipartite network to a one-mode network by defining the weights as the number of co-occurrences. In other cases, such networks are created directly from raw count data. For instance, in a physical co-location network, we might create an edge between two individuals, where the edge weight indicates the number of times the individuals were in the same place at the same time. Network analysis techniques have been used to explore co-occurrence patterns in a range of domains.  One common application area has been in bibliometrics, where co-citation networks have been analysed \cite{gmur03invisible}. Here a co-citation link exists between two research papers if they are both cited by a third paper. By analysing the structure of such networks, it is possible to map the research activities, collaborations, and trends within and across research fields \cite{white1998visualizing}. Other common examples of the use of co-occurrence networks include the analysis of word co-occurrence patterns in natural language processing \cite{edmonds97word}, the analysis of co-purchasing trends in online retail systems \cite{ghoshal11nature}, and the use of co-listed information to recommend users to follow in the context of online social media platforms \cite{greene2012aggregating}.

\subsection{Community Finding}

When analysing networks in many domains, we will often be interested in performing community detection, where the goal is to identify the underlying group structures in the data. Typically this is performed as an unsupervised task. A substantial amount of work in this area has focused on the detection of disjoint communities, where each node belongs to at most one community. Algorithms in this context can be broadly grouped into three types: 

\noindent\textit{(1) Hierarchical algorithms} construct a tree of communities based on the network topology. These can be one of two types: divisive algorithms \cite{girvan2002community} or agglomerative algorithms \cite{clauset2004finding}. \textit{(2) Modularity-based algorithms} optimise the well-known modularity objective function to uncover communities in a network \cite{newman2006modularity}. \textit{(3) Other algorithms} which include those based on label propagation approaches \cite{xie11slpa}, spectral methods that make use of the eigenvectors of a Laplacian  or standard matrix, and methods based on statistical modelling \cite{fortunato10review}.

In many real-world networks, we observe ``pervasive overlap'', such that individuals frequently belong to many highly-overlapping communities \cite{ahn10link}. Therefore, overlapping community finding algorithms \cite{amelio2014overlapping} have been developed for application to these networks. These can be classified into four main categories:

\noindent\textit{(1) Node seeding and local expansion algorithms} detect communities by starting from a node, or a small group of nodes, and then expanding these into a community using some fitness function. OSLOM \cite{lancichinetti11oslom} is one example, that expands communities based on a fitness function measuring the statistical significance of communities with respect to random variations. We discuss this algorithm in more detail in Section \ref{sec:methods-comm}.
\textit{(2) Clique expansion methods} use a group of fully-connected nodes, called a clique, as the starting point for building larger communities. CFinder \cite{adamcsek2006cfinder} and Greedy Clique Expansion (GCE) \cite{lee10gce} are examples of this type of algorithm. \textit{(3) Link clustering algorithms} detect communities by splitting the network edges rather than the nodes. GA-NET+ \cite{pizzuti2009overlapped} is a good example of this category. \textit{(4) Label propagation algorithms} attempt to group each node into a community based on its neighbouring nodes' affinities. An example of this approach is the Speaker Label Propagation Algorithm (SLPA) \cite{xie11slpa}.

\section{Building the Dublin Meetup Network}
\label{sec:methods}

This section describes how data was collected using the meetup.com API, how a meetup co-occurrence network was constructed, the application of community finding methods to this network, and the use of text analytics methods to label and explain communities found in the network. 

\subsection{Data Collection}
\label{sec:methods-collect}
The Meetup.com website provides an open API\footnote{Meetup.com API: \url{https://www.meetup.com/meetup_api/}}
 that allows access to data from the meetup.com platform. Building a meetup co-occurrence network required data about meetup groups in Dublin and the users that are members of each group. The \emph{/find/groups} API call generates a list of all meetups in a specified country. Metadata providing the name, description, and host city of each meetup is also returned. Using this call we generated a list of all meetups in Ireland and then filtered this to exclude those not hosted in Dublin. Private meetups were also excluded. This left a filtered set of 1,482 Dublin-based public meetups. The \emph{/2/members} API call generates a list of member IDs for all members of a specific meetup group. We used this call to generate a list of all of the members of each Dublin-based public meetup group identified in the previous step. 

\subsection{Network Construction}
\label{sec:methods-con}

\begin{figure}[bt]
\centering
\includegraphics[width=0.4\linewidth]{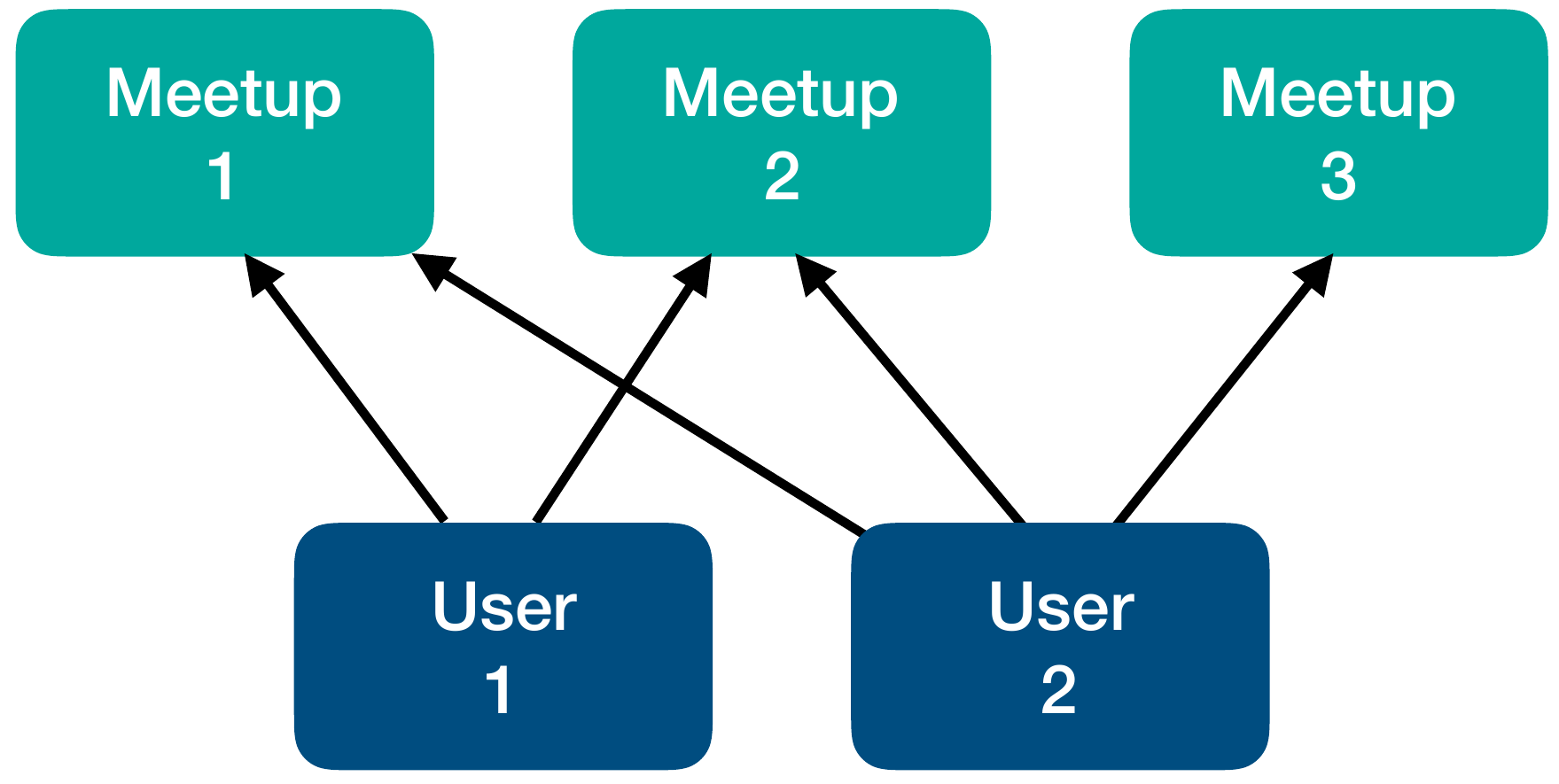}
\caption{Example of a co-membership relationship between meetups, where an arrow indicates that a user is a member of a meetup. 
}
\label{fig:example}
\end{figure}

Motivated by the concept of co-citation networks in bibliometrics, we create an undirected weighted graph based on the member overlaps, or \emph{co-memberships}, between pairs of meetups. \reffig{fig:example} illustrates this approach. As \emph{Meetup 1} and \emph{Meetup 2} share common users, a link would exist between them in a co-occurrence graph, whereas neither of these would link to \emph{Meetup 3} as no common users exist. 

In the proposed network, each unique meetup is represented as a node, and a weighted edge between two nodes represents an association between the two meetups represented by its endpoints. Every meetup has a set of registered members, which allows us to measure the degree of overlap between the member sets for pairs of meetups. Rather than looking at the raw number of common members, we use a normalised value. Specifically, the degree of association between two meetups is calculated using the Jaccard index \cite{jaccard12index}, which has been previously used in co-citation analysis \cite{small1973co}. Formally, we calculate the edge weight $w_{ij}$  between meetups $i$ and $j$, with sets of members $M_i$ and $M_j$ respectively, as:
\begin{equation}
w_{ij} = \frac{|M_i \cap M_j|}{|M_i \cup M_j|}
\end{equation}
Edges only exist between pairs of nodes for which $w_{ij}>0$. We refer to the resulting network as the \emph{normalised meetup co-membership network} or more simply the \emph{Dublin meetup network}.




\subsection{Finding Communities}
\label{sec:methods-comm}

We apply community finding to the Dublin meetup network to organise it into a smaller number of communities of related meetups that are easier to interpret, as opposed to manually inspecting a large number of meetups individually. Given that we might expect some users to be members of a diverse set of meetups,  we apply an overlapping community finding approach  to the co-membership network, which allows each meetup to potentially belong to multiple communities. Specifically, we apply the popular OSLOM algorithm \cite{lancichinetti11oslom}, which is suitable for use on weighted networks. 

OSLOM follows a greedy expansion strategy to detect communities by optimising a local fitness function on seed nodes based on statistical measurement. This is done in three stages. In the first stage, the algorithm searches for significant clusters by selecting a node at random as a seed then expanding it into a larger community. During this expansion, the algorithm tests the fitness score of the neighbouring nodes of the seed node by using a statistical test that evaluates the significance of each neighbouring node to be added to the seed. In the second stage, for each community, the algorithm performs a process which involves removing or adding nodes to maximise its significance score. In the third stage, the hierarchical levels of the communities are detected. Due to the presence of a stochastic element during calculating a node's significant score the above stages are repeated several times and aggregated to create a stable set of communities.  

To identify communities, we applied the undirected version of OSLOM  to the meetup network using a range of values  $[0.01,0.5]$ for the resolution parameter, which indirectly controls the size of the communities identified by the algorithm. After each run, we filtered communities containing $< 5$ nodes, on the basis that these do not represent significant groupings of meetups. Based on manual inspection, a value of $0.1$ for the resolution parameter provided a good trade-off between ensuring that communities were coherent, while also ensuring that the number of duplicate communities (\ie related to identical topics) was minimised. 

\subsection{Labelling Communities}
\label{sec:methods-labelling}
To extract meaningful insights from the community finding results, and to examine the topical coherence of these communities, we produce human-interpretable labels for each community. This allows us to explain and understand the groups at a high level. Fortunately, the meetup.com API provides a rich set of metadata  which can be used for this purpose. Specifically, we propose a custom approach for labelling each community based on the short name field and the longer description field associated each meetup assigned to that community. Formally, we generate \emph{name labels} for communities as follows: 
\begin{enumerate}
    \item For each meetup name field, extract all alphanumeric terms and filter out common stopwords (\eg ``the'', ``meetup''', ``group'').
    \item Construct a meetup-term matrix $A$, such that each row corresponds to a meetup, each column corresponds to a term, and each entry indicates the number of times a term appears in a meetup name.
    \item Apply standard log-based TF-IDF weighting to the matrix, and normalise the rows to unit length to account for different name lengths.
    \item For each community $C$:
    \begin{enumerate}
    \item From $A$, compute the mean vector of the rows corresponding to the meetups which have been assigned to $C$.
    \item Rank the values in the mean vector in descending order, and select the top $t$ terms to create a name label.
    \end{enumerate}
\end{enumerate}
We apply an analogous procedure to generate \emph{description labels} for communities, based on the longer meetup description text.

\section{Exploring the Dublin Meetup Network}
\label{sec:eval}

In the original co-membership graph, each meetup group is represented by a node and the edges indicates the weight of the connection between pairs of meetups. The network consists of 1,482 nodes connected by 1,416,326 weighted edges, which represents a single connected component.


\subsection{Network Characterisation}
\label{sec:methods-char}


\begin{table}[!b]
\centering
\begin{subfigure}[t]{\textwidth}
\centering
\scriptsize{
\begin{tabular}{|r|p{6.5cm}|r|}
\hline
\textbf{Rank} & \textbf{Meetup Name}                                 & \textbf{Score} \\\hline
1             & Dublin Artificial Intelligence \& Deep Learning & 0.0659         \\
2             & Big Data Developers in Dublin                   & 0.0658         \\
3             & Data Scientists Ireland                         & 0.0654         \\
4             & Zalando Tech Events Dublin                      & 0.0651         \\
5             & Machine Learning Dublin                         & 0.0640         \\
6             & Data Science and Engineering Club               & 0.0639         \\
7             & Hackers and Founders Dublin                     & 0.0638         \\
8             & GDG Dublin                                      & 0.0635         \\
9             & Dublin Startup Founder 101                      & 0.0632         \\
10            & Dublin - Coder Forge                            & 0.0630    \\\hline    
\end{tabular}
}
\caption{Weighted eigenvector centrality}\label{tab:centrality_wec}
\end{subfigure}
\begin{subfigure}[t]{\textwidth}
\centering
\scriptsize{
\begin{tabular}{|r|p{6.5cm}|r|}
\hline
\textbf{Rank} & \textbf{Meetup Name}                                    & \textbf{Score} \\\hline
1             & Speak English Dublin                               & 0.1053         \\
2             & AWS User Group Dublin                              & 0.0467         \\
3             & New and Not So New In Dublin                       & 0.0428         \\
4             & English Language Conversation with English Teacher & 0.0349         \\
5             & Events, Drinks and Dancing in Dublin               & 0.0327         \\
6             & Python Ireland                                     & 0.0323         \\
7             & Dublin UX                                          & 0.0289         \\
8             & Dublin Indoor Football at The Soccer Dome          & 0.0260         \\
9             & Dublin LGBT Social Meetup                          & 0.0229         \\
10            & Socializing Dublin                              & 0.0218 \\\hline       
\end{tabular}
}
\caption{Weighted betweenness centrality}\label{tab:centrality_wbc}
\end{subfigure}
\vskip 0.5em
\caption{Top 10 most central meetups in the Dublin meetup network, as ranked by (a) weighted eigenvector centrality and (b) weighted betweenness centrality.}
\label{tab:centrality}
\end{table}


As is typical of many co-occurrence networks, the Dublin meetup network is highly dense, with edges present between 64.5\% of all possible pairs of nodes. However, there is some variation in terms of the weights on these edges. We observe that 86.63\% of the weighted edges have values $\leq 0.01$. This is indicative of a relatively small intersection between the memberships of many meetups.

When ranked by number of members, unsurprisingly, the three largest meetups are from the social sphere: \emph{New and Not So New In Dublin} (21,149 members), \emph{Events, Drinks and Dancing in Dublin} (16,582 members) and \emph{Dublin International} (14,812 members). It is more informative, however, to measure the importance of meetup nodes in the overall meetup network and we do this using centrality analysis. A range of measures have previously been proposed for this task. We focus on two popular measures of centrality which are designed for use on weighted networks:
\begin{enumerate}
    \item In \emph{weighted eigenvector centrality}, a node is deemed more important if it is connected to other important nodes. In the weighted variant of this measure, centrality scores are calculated based on the first left eigenvector of the weighted graph adjacency matrix  \cite{bonacich1987power}.
    \item The \emph{weighted betweenness centrality} measure identifies strategic bridges in a network. Nodes that occur on many shortest paths between other nodes in the network have high centrality. In the weighted variant of betweenness, the weighted distances between nodes are taken into account \cite{brandes2008variants}.
\end{enumerate}

The top 10 meetups as ranked by each measure are listed in \reftab{tab:centrality}. It is interesting to note the significant difference between the lists generated by these two approaches---in fact there are no meetups that occur in both lists. The dominance of technology related meetups in \reftab{tab:centrality_wec} reflects the vibrance of the tech community in Ireland and indicates a slight bias on the meetup.com platform towards technology savvy users. It also suggests the existence of a cluster of similar meetups attended by a core group of overlapping members. The meetups in \reftab{tab:centrality_wbc} reflect the fact that betweenness centrality measures the ability of nodes in a network to connect disparate parts of that network. Although there are some technology related meetups here, most focus on topics of broad appeal (\eg \emph{Speak English Dublin}) that are likely to attract members with disparate other interests. 

\begin{figure}[!t]
\centering
    \centering
    \includegraphics[width=0.8\linewidth]{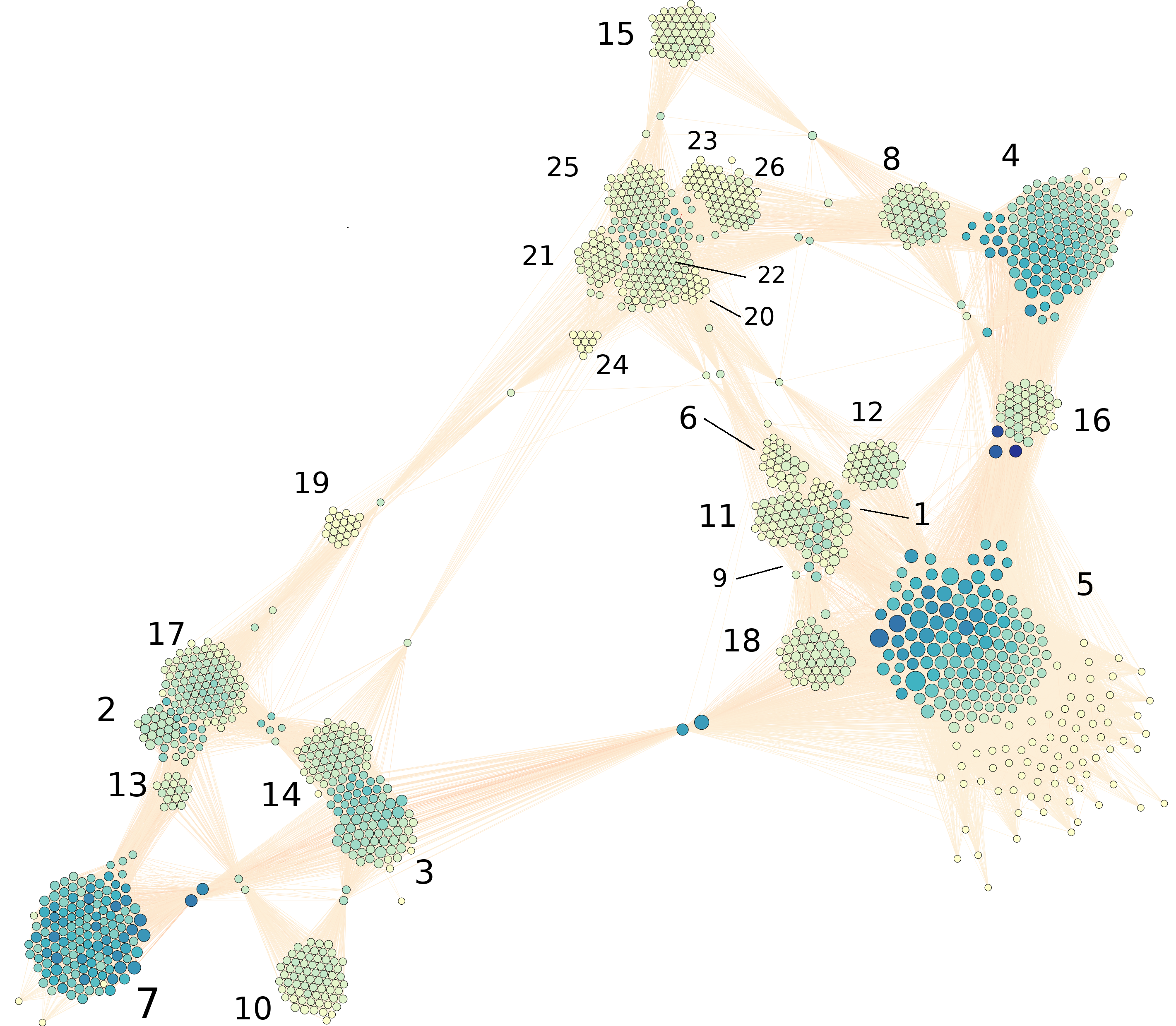}
    \caption{A visualisation of the meetup network for Dublin, where only intra-community edges are shown. Each node represents a meetup, where its size is proportional to the number of members in the meetup. Darker colour indicate a higher weighted degree of the nodes, numbers indicate community Ids.}
    \label{fig:meetup_graph_full}
\end{figure}

\subsection{Exploring Communities}
The application of the OSLOM community finding algorithm yielded a model with 26 communities, ranging in size from 17 to 216 meetups. The mean community size was 65. \reftab{tab:comm} summarises details for the 15 largest communities identified in the normalised meetup co-membership network (full list is available in \footref{note1}). For each community, we report its size (\ie number of meetups) and labels generated based on meetup name and description metadata, using the approach described in \refsec{sec:methods-comm}. These labels indicate a diverse range of communities, covering topical areas such as technology, travel, self-help, music, and entrepreneurship. Despite the fact that the name and description fields typically differ considerably in length, it is interesting to note the relatively high level of overlap for the terms appearing on both lists for the same community.

\begin{table}[!tb]
\centering
\scriptsize{
\begin{tabular}{|r|r|p{4.9cm}|p{4.9cm}|}
\hline\textbf{Id} & \textbf{Size} & \textbf{Name Label}                                                                                   & \textbf{Description Label}                                                                             \\\hline
5  & 216  & hiking, international, wicklow, friends, yoga, book, culture, adventure, language, travel             & fun, members, friends, time, hikes, free, social, friendly, looking, food                              \\
4  & 148  & meditation, yoga, healing, spiritual, heart, sound, empowerment, soul, life, positive                 & healing, life, meditation, experience, self, energy, practice, spiritual, mind, mindfulness            \\
7  & 137  & data, user, science, tech, engineering, big, cloud, users, things, learning                           & data, programming, developers, community, code, science, software, technology, technologies, learn     \\
17 & 118  & user, tech, security, cloud, sharepoint, technology, game, software, data, crypto                     & data, learn, share, learning, developers, cloud, community, security, technology, software             \\
14 & 84   & business, digital, marketing, startup, entrepreneurs, network, job, professionals, innovation, market & business, marketing, digital, entrepreneurs, startup, market, network, owners, sales, job              \\
22 & 80   & yoga, meditation, workshop, stress, dun, laoghaire, camino, running, dance, therapy                   & yoga, life, body, meditation, class, health, classes, practice, energy, mind                           \\
3  & 78   & startup, entrepreneurs, digital, lean, business, marketing, agile, growth, product, innovation        & business, entrepreneurs, marketing, startup, networking, digital, lean, product, community, innovation \\
25 & 77   & yoga, health, happiness, meditation, vegan, prayer, empowerment, circle, centre, self                 & yoga, life, meditation, help, support, healing, learn, world, health, work                             \\
10 & 71   & user, mysql, traders, developers, tech, js, product, data, sprint, net                                & learn, product, developers, mysql, share, community, meetups, professionals, technologies, engineers   \\
18 & 63   & music, singles, rock, social, travel, south, international, fans, electronic, 30s                     & music, night, friends, fun, singles, rock, singing, love, members, sing                                \\
8  & 61   & yoga, meditation, health, healing, classes, relaxation, self, body, light, sound                      & yoga, meditation, body, classes, life, mind, healing, health, practice, nature                         \\
21 & 54   & empowerment, self, book, support, health, workshop, eating, therapy, life, development                & life, world, diet, work, feel, learn, share, spiritual, ideas, find                                    \\
15 & 53   & circle, things, drinks, city, fun, hike, ladies, social, friends, book                                & drinks, friends, women, fun, book, food, wants, single, cinema, dinner                                 \\
16 & 53   & dance, dancing, yoga, classes, movement, salsa, fitness, class, set, handstand                        & dance, classes, dancing, fun, fitness, workout, 8pm, levels, class, movement                           \\
26 & 52   & soul, prayer, network, life, healing, workshop, empowerment, biodanza, centre, body                   & life, god, healing, faith, spiritual, love, evening, work, reiki, chat\\\hline                                
\end{tabular}
}
\vskip 0.5em
\caption{Summary of the 15 largest communities identified in the Dublin meetup network. Community labels are generated based on the top 10 terms appearing in the meetup name and description fields.}
\label{tab:comm}
\end{table}

To further explore the results produced by OSLOM, for each community we constructed a subgraph of the original network, and then ranked the nodes assigned to their community based on their centrality within that induced subgraph, as calculated by \emph{weighted degree centrality}. The score for a node $i$ is defined as the sum of the weights of the edges connecting $i$ and its neighbours. \reftab{tab:nodes} lists the top 3 most central meetups in each of the 15 largest communities.

These 26 subgraphs are shown in \reffig{fig:meetup_graph_full} generated using the  MultiGravity Force Atlas 2 graph layout algorithm provided by the Gephi tool \cite{gephi2009}. The connections between these subgraphs are due to the overlapping meetups between the communities, which makes the entire meetup.com community a single connected component. The size of each node in \reffig{fig:meetup_graph_full} is proportional to the number of members in that group; the colour of each node indicates its weighted degree.

\begin{figure}[!t]
\centering
    \includegraphics[width=0.9\linewidth]{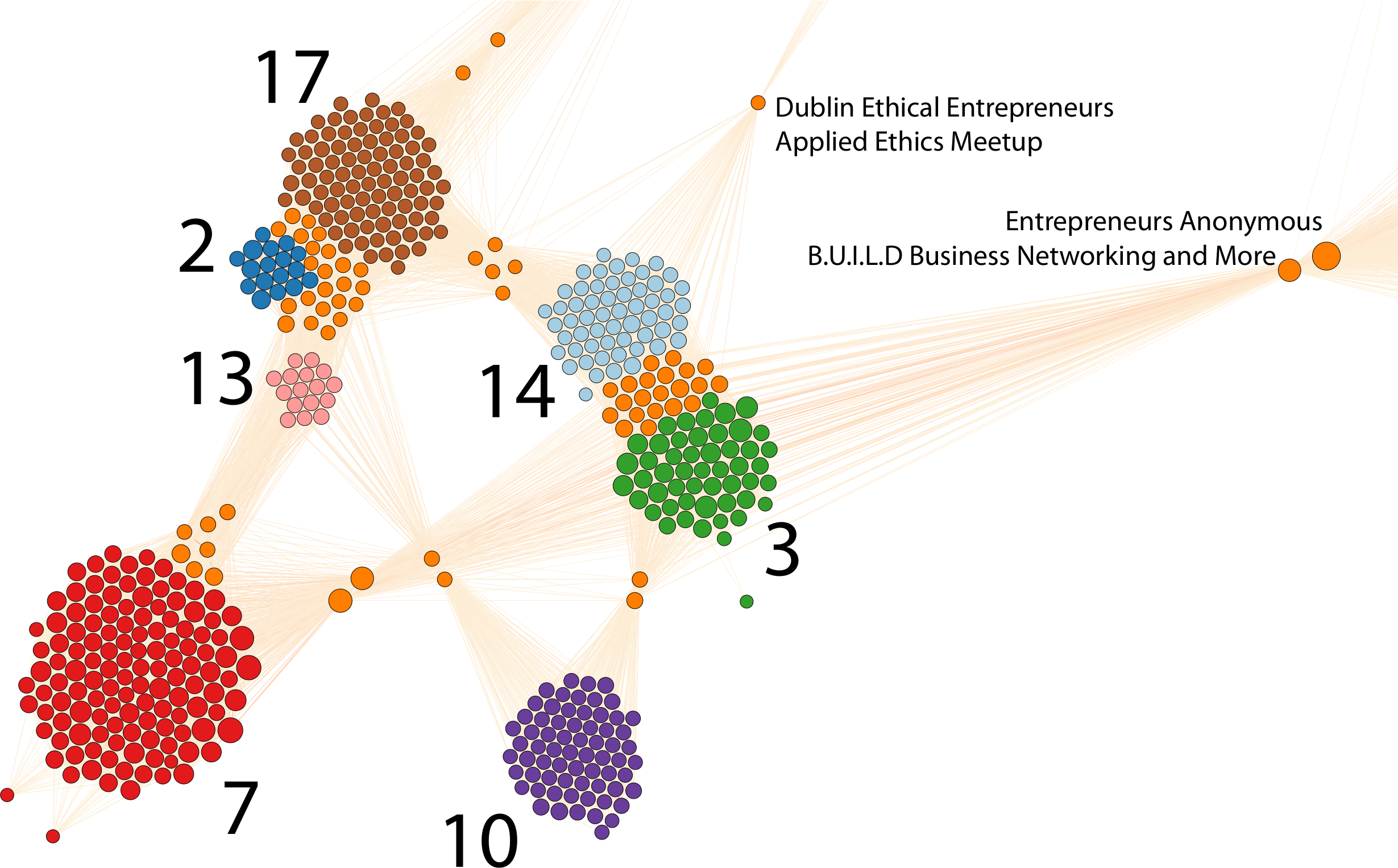}
    \vskip -0.2em
    \caption{The communities of technology focused  meetups, each highlighted in a unique colour. Meetups in multiple communities are highlighted in orange. }\label{fig:tech_meetup_comm_graph}
\end{figure}

\begin{table}[tb]
\centering
\scriptsize{
\begin{tabular}{|r|r|p{9.6cm}|}
\hline\textbf{Id} & \textbf{\% Over} & \textbf{Top 3 Meetups}\\\hline
5           & 16.7             & Kick Ass Adventures, Awesome Daytrips, Live Music  Artistic and Cultural Events \\
4           & 12.2             & Dublin Meditation Meetup, Dublin Self-Empowerment Coaching Meetup, Holistic Events Dublin \\
7           & 6.6              & Data Science and Engineering Club, Dublin Node.js Meetup, Dublin Artificial Intelligence \& Deep Learning \\
17          & 24.6             & Social Software Development Meetup in Dublin, Amazic Ireland: Enterprise Docker and DevOps, Fairy Tale Artificial General Intelligence Solutions \\
14          & 32.1             & Idea 2 Scale Dublin, SEO Strategy, Business Improvement \& Innovation Network \\
22          & 35.0               & Ki Yoga Massage, Dublin Wellness Meetup, Practical Ayurveda \\
3           & 34.6             & Startup Grind Dublin, Dublin Startup Founder 101, Start-Ed Dublin: Free legal forum for Startups \\
25          & 35.1             & Radical Forgiveness, Ki Yoga Massage, Dublin Grief Support Group \\
10          & 5.6              & WebRTC Meetup Ireland, Serverless-Dublin, NetDevOps Dublin \\
18          & 23.8             & Dublin Singles Party Meetup, Single Professionals Dublin, Dublin Nightlife \\
8           & 23.0               & Holistic Health and Wellness, Gentle Yoga, Beginner's Meditation Course \\
21          & 42.6             & Radical Forgiveness, Radical Human Expression, Dublin YouSA Meetup \\
15          & 5.7              & Finding things to do in Dublin, 40 (or almost) and fabulous, Random things to do in Dublin \\
16          & 26.4             & Stefano Argentine Tango, Samba Gafiera/Brazilian Zouk Dance Meetup, Dublin Dancing Meetup \\
26          & 25.0               & Healingoptions4u, Radical Forgiveness, Pilgrims' Prayer\\\hline               
\end{tabular}
}
\vskip 0.7em
\caption{The top 3 most central meetups in each of the 15 largest communities found. \emph{\% Over} shows the percentage of overlapping nodes in each community.}
\label{tab:nodes}
\end{table}

The level of overlap between communities was not as high as might be expected. In total, 197 of the 1,482 meetups were assigned to more than one community. Of these, 176 appear in two communities, 20 in three communities, and a single meetup appears in four communities (\emph{Headless Awareness Dublin}). \reftab{tab:nodes} also reports the percentage of overlapping nodes in each of the 15 largest communities---\ie how many nodes assigned to the community were also assigned to at least one other community. For some communities, a reasonably high proportion of nodes are overlapping (\eg communities 21, 22, 25 in \reftab{tab:nodes}), while in other cases the vast majority of nodes belong exclusively to that community (\eg communities 7, 10, 15 in \reftab{tab:nodes}).

To further illustrate the value of community finding, we present a more detailed analysis of a small subset of seven of the 26 communities found that have been manually identified as relating to technology. This subset is highlighted in \reffig{fig:tech_meetup_comm_graph}, in which each community has a unique colour and meetups belonging to multiple communities are highlighted in orange. The community IDs from Tables \ref{tab:comm} and \ref{tab:nodes} are also shown. 

The biggest community in this subset (highlighted in red in \reffig{fig:tech_meetup_comm_graph}) is community 7. \emph{Data Science and Engineering Club} with $1,886$ members has the highest weighted degree within this community, and shares  more than 10\% of its members with 19 other meetups in the community. The top five meetups with which the \emph{Data Science and Engineering Club} meetup shares its members are: \emph{Dublin Data Science} ($30.8\%$), \emph{Dublin Artificial Intelligence \& Deep Learning} ($22.1$\%), \emph{Data Scientists Ireland} ($22.0$\%), \emph{Machine Learning Dublin} ($21.0$\%) and \emph{PyData Dublin} ($19.1$\%). This shows a strong concentration around data science and machine learning. We see a similar concentration in community 2 (dark blue in \reffig{fig:tech_meetup_comm_graph}) which contains meetups largely in the areas of cryptography and blockchain. 
The meetups in this community also share significant membership with community 17 (brown in \reffig{fig:tech_meetup_comm_graph}).

It is interesting that this subset of the meetup community structure is connected to the remainder of the network largely by a small number of meetups. In particular \emph{Entrepreneurs Anonymous Dublin}, \emph{B.U.I.L.D business networking and more Dublin}, and \emph{Dublin Ethical Entrepreneurs Applied Ethics Meetup} mediate the majority of the links of the technology communities to the rest of the Dublin meetup network. These nodes are also members of other communities.

Outside of technology there are other small, interesting, well defined social communities. For example, community 12 focuses broadly on art, the meetups in community 1 are largely focused on LGBTQ+ interests, community 9 contains meetups organised around playing board games, and community 6 is largely related to sports. The overlaps between these communities are also interesting. For example, community 12 has a relatively large overlap with community 1, and community 9 has a relatively large overlap with community 6.

\section{Conclusions}
\label{sec:conclusions}

This paper demonstrated the use of network analysis techniques to explore public data collected from the meetup.com platform,
to characterise the fabric of a city.  A co-membership network of meetups from Dublin, Ireland was constructed. This network allowed us to reveal the most important meetups in Dublin via measures of node centrality. To uncover the structure of the network at a higher level, community finding techniques were applied. By subsequently applying text analysis procedures to the aggregated metadata associated with each community, we have shown that thematically-coherent communities exist within Dublin's Meetup ecosphere. We also observed a limited degree of overlap between certain communities, where users might naturally share common interests.

Although the present study specifically focuses on Dublin's meetup network, using the same framework, the underlying communities of other cities could also be explored, and future work will develop a tool to support this. It would also be interesting to incorporate additional layers of meetup metadata into the network construction process. For example, some meetup groups are much more active than others, as some users are much more active than others, and this could be used to filter nodes in the network, and to influence the weight of edges. 


\bibliographystyle{splncs03}
\bibliography{meetup_network_dublin} 

\end{document}